\documentclass[12pt]{article}

\usepackage{amsmath,amssymb}
\usepackage{graphicx}

\newcommand{\beq}{\begin{equation}}
\newcommand{\eeq}{\end{equation}}
\newcommand{\bea}{\begin{eqnarray}}
\newcommand{\eea}{\end{eqnarray}}

\newcommand{\Tr}{\, \mathrm{Tr}}
\newcommand{\he}{\hat{\epsilon}}
\newcommand{\integral}{\int d^2 x \sqrt{g}} 

\begin{document}

\pagestyle{empty}
\begin{center}

\vspace*{30mm}
{\LARGE Canonical coordinates for\\ \vspace*{3mm} Yang-Mills-Chern-Simons theory}

\vspace*{30mm}
{\large M{\aa}ns Henningson}

\vspace*{5mm}
Department of Physics\\
University of Gothenburg\\
G\"oteborg, Sweden\\[3mm]
{\tt mans.henningson@physics.gu.se}

\vspace*{25mm}{\bf Abstract:}
\end{center}
We consider the classical field theory of 2+1-dimensional Yang-Mills-Chern-Simons theory on an arbitrary spatial manifold. We first define a gauge covariant transverse electric field strength, which together with the gauge covariant scalar magnetic field strength can be taken as coordinates on the classical phase space. We then determine the Poisson-Dirac bracket and find that these coordinates are canonically conjugate to each other. The Hamiltonian is non-polynomial when expressed in terms of these coordinates, but can be expanded in a power series in the coupling constant with polynomial coefficients.

\newpage \pagestyle{plain}

\section{Introduction}
Three-dimensional Yang-Mills theory \cite{Yang-Mills} with a Chern-Simons interaction \cite{Chern-Simons} is governed by the Lagrangian density \cite{Schonfeld}\cite{Jackiw-Templeton}\cite{Deser-Jackiw-Templeton}\footnote{Here $\lambda$ is the coupling constant (of dimension $\text{mass}^{1/2}$) and $k$ is an integer level. Our conventions for the gauge covariant derivate $D_\mu$ of a field $X$ in the adjoint representation and the curvature $F_{\mu \nu}$ of the Yang-Mills connection $A_\mu$ are $D_\mu X = \nabla_\mu X + i [X, A_\mu]$ and $F_{\mu \nu} = \nabla_\mu A_\nu - \nabla_\nu A_\mu - i [A_\mu, A_\nu]$, with $\nabla_\mu$ being the covariant derivative constructed out of the spin-connection on the space-time manifold. This gives the commutator of gauge covariant derivatives $[D_\mu, D_\nu] X = i [X, F_{\mu \nu}]$. These conventions are compatible with $X$, $A_\mu$ and $F_{\mu \nu}$ all being Hermitian. With an $SU (N)$ gauge group, $\Tr$ is the trace in the fundamental $N$-dimensional representation.}
 
\beq \label{Lagrangian density}
{\cal L} = \frac{1}{8 \lambda^2} \Tr \left(F^{\mu \nu} F_{\mu \nu} \right) + \frac{k}{4 \pi} \frac{1}{\sqrt{G}} \epsilon^{\mu \nu \rho} \Tr \left(A_\mu \partial_\nu A_\rho - \frac{2 i}{3} A_\mu A_\nu A_\rho \right) .
\eeq

The Bianchi identity and the equations of motion read
\bea \label{EOM_Bianchi}
D_\mu F_{\nu \rho} + D_\nu F_{\rho \mu} + D_\rho F_{\mu \nu} & = & 0 \cr
\frac{1}{2 \lambda^2} D_\nu F^{\mu \nu} + \frac{k}{4 \pi} \epsilon^{\mu \nu \rho} F_{\nu \rho} & = & 0 . 
\eea
It follows from these equations that a current of the form $J^\mu = T^{\mu \nu} K_\nu$, where 
\beq
T^{\mu \nu} = \frac{1}{2 \lambda^2} \Tr \left(F^{\mu \rho} F^\nu{}_\rho - \frac{1}{4} G^{\mu \nu} F^{\rho \sigma} F_{\rho \sigma} \right)
\eeq
is the energy momentum tensor and the vector field $K_\mu$ obeys the Killing condition $\nabla_\mu K_\nu + \nabla_\nu K_\mu = 0$, is conserved, i.e. $\nabla_\mu J^\mu = 0$.

In this note, we will be concerned with the case where the space-time metric $G_{\mu \nu}$ admits a time-like Killing vector field, which by a choice of space-time coordinates $x^\mu$, $\mu = 0, 1, 2$ can be taken as $K_0 = 1$, $K_i = 0$, $i = 1, 2$.\footnote{The spatial metric $g_{i j}$ may admits further Killing vector fields, that would give rise to additional conserved currents.} The space-time metric then takes the form $G_{\mu \nu} d x^\mu d x^\nu = - (d t)^2 + g_{i j} d x^i d x^j$, and the equations (\ref{EOM_Bianchi}) can be rewritten in a $2+1$-dimensional notation:
\bea \label{EOM}
D_0 B - \he^{i j} D_i E_j & = & 0 \cr
\frac{1}{2 \lambda^2} (D_0 E^i + \he^{i j} D_j B) - \frac{k}{2 \pi} \he^{i j} E_j & = & 0
\eea
and
\bea \label{constraint}
- \frac{1}{2 \lambda^2} D_i E^i + \frac{k}{2 \pi} B & = & 0 ,
\eea 
where 
\bea
B & = & \frac{1}{2} \he^{i j} F_{i j} \cr
E_i & = & F_{0 i} 
\eea
are the magnetic and electric field strengths respectively.\footnote{We define the {\it tensor} $\he^{i j} = \frac{1}{\sqrt{g}} \epsilon^{i j}$ in terms of the Levi-Civita tensor density $\epsilon^{i j}$.} The equations (\ref{EOM}) determine the time development of $B$ and $E^i$ respectively, while equation (\ref{constraint}) is a constraint that needs to be imposed on the data at some initial time. The time component of the current constructed from the time-like Killing vector is the Hamiltonian density
\bea \label{Hamiltonian_density}
{\cal H} & = &T^{0 0} \cr
& = & \frac{1}{2 \lambda^2} \Tr (E_i E^i + B B) ,
\eea
the spatial integral of which is the Hamiltonian
\beq \label{Hamiltonian}
H = \integral {\cal H} .
\eeq

The phase space of the theory is given by the space of solutions to the equations (\ref{EOM}) and (\ref{constraint}) modulo (infinitesimal) gauge transformations acting as
\bea \label{gauge transformation}
\delta A_\mu & = & D_\mu \Lambda \cr
\delta B & = & i [B, \Lambda] \cr
\delta E_i & = & i [E_i, \Lambda] 
\eea
with a Lie algebra-valued parameter $\Lambda$. The phase space is endowed with a symplectic structure, i.e. a closed, non-degenerate two-form given by a Poisson-Dirac bracket $\{ . , . \}$. The equations of motion (\ref{EOM}) can then be cast in Hamiltonian form:
\bea \label{HamiltonianEOM}
\{ H, B \} & = & D_0 B + i [B, \Lambda] \cr
& = & \he^{i j} D_i E_j +  i [B, \Lambda] \cr
\{ H, E^i \}  & = & D_0 E^i + i [E^i, \Lambda] \cr
& = & - \he^{i j} D_j B + \frac{\lambda^2 k}{\pi} \he^{i j} E_j + i [E^i, \Lambda] ,
\eea
where we have included a possible infinitesimal gauge transformation with parameter $\Lambda$.\footnote{The need to include such a gauge transformation is related to the fact that the Chern-Simons term Lagrangian density (\ref{Lagrangian density}) changes by a total derivative under a gauge transformation (\ref{gauge transformation}): $\delta {\cal L} = \frac{k}{4 \pi} \frac{1}{\sqrt{G}} \epsilon^{\mu \nu \rho} D_\mu (\Lambda F_{\nu \rho})$. Of course, the time development of gauge invariant physical observables are not affected in this way.}

The aim of this note is to describe a convenient set of coordinates to parametrise this phase space. We will then compute the symplectic structure by giving the Poisson-Dirac bracket between these coordinates. We will see that our coordinates are canonically conjugate to each other, i.e. they are Darboux coordinates. Finally, we will express the Hamiltonian in terms of these coordinates. The Hamiltonian is non-polynomial, but can be expanded in a power-series in the coupling constant $\lambda$ with coefficients that are polynomial in the (rescaled) coordinates.

There is a vast literature on 2+1-dimensional Yang-Mills theory, in particular using a covariant path-integral formalism. An early reference in a Hamiltonian framework is \cite{Feynman}. The use of gauge covariant variables, e.g. Wilson loops \cite{Wilson} or the field strength \cite{Goldstone-Jackiw}\cite{Halpern}, instead of the gauge potential as dynamical variables also has a long history. The combination of these approaches in a formulation of Hamiltonian lattice gauge theory was pioneered in \cite{Kogut-Susskind}. An approach, rather unrelated to ours, of constructing a set of phase space variables was initiated in \cite{Karabali-Nair-Kim}. (See \cite{Nair} for a review, which includes many further references). However, although the main ideas entering into our constructions are well known, we are not aware of any previous description of these gauge covariant and canonical phase space coordinates in the literature. We hope that the may prove useful for the purpose of canonical quantization of Yang-Mills-Chern-Simons theory.

\section{The coordinates}
The electric field strength $E_i$ may, as any vector field in two dimensions, be decomposed into its longitudinal and transverse scalar components $E_\parallel$ and $E_\perp$:\footnote{In this note, we will not be concerned with any issues connected to possible zero-modes.}
\beq \label{parallel_transverse}
E_i = D_i E_\parallel + \he_i{}^j D_j E_\perp .
\eeq
In principle, this equation may be inverted to express the gauge covariant components $E_\parallel$ and $E_\perp$ in terms of the vector components $E_i$, but not in closed form.
 
The constraint (\ref{constraint}) can now be written as\footnote{We define $D^2 = D_i D^i$ with inverse $D^{-2}$, and similarly $\nabla^2 = \nabla_i \nabla^i$ with inverse $\nabla^{-2}$, again not worrying about possible zero-modes.} \footnote{By the identity $\he^{i j} D_i D_j E_\perp = i [E_\perp, B]$.}
\beq \label{newconstraint}
- \frac{1}{2 \lambda^2} \left(D^2 E_\parallel + i [E_\perp, B] \right) + \frac{k}{2 \pi} B = 0 ,
\eeq
and can be solved by expressing $E_\parallel$ in terms of $E_\perp$ and the connection $A_i$:
\beq \label{E_parallel}
E_\parallel = D^{-2} \left(-i [E_\perp, B] + \frac{\lambda^2 k}{\pi} B \right) .
\eeq

We will now take the transverse electric field $E_\perp$ together with the magnetic field $B$ as our phase space coordinates. They are gauge covariant, but of course not gauge invariant, so to determine a point in the phase space, they must be supplemented by a gauge condition
\beq \label{gaugechoice}
\phi_A = 0 .
\eeq
Here the Lie algebra-valued scalar field $\phi_A$ is some fairly arbitrary (linear) functional of the spatial gauge connection $A_i$ chosen so that this equation determines a single point on each orbit of the group of gauge transformations. This allows us to express $A_i$ in terms of $B$. For many purposes, it will e.g. be convenient to use the Coulomb gauge condition $\phi_A = \nabla_i A^i = 0$.  The general solution is then $A_i = \epsilon_i{}^j \nabla_j A$, where $A$ is a Lie algebra-valued scalar field. This gives $B = - \nabla^2 A - \frac{i}{2} \he^{i j} [\nabla_i A, \nabla_j A]$, which may be inverted to express $A$, and thus $A_i$, as a power series in $B$:
\beq \label{Coulomb}
A_i = \epsilon_i{}^j \nabla_j \left(- \nabla^{-2} B - \frac{i}{2} \he^{k l} \nabla^{-2} [\nabla_k \nabla^{-2} B, \nabla_l \nabla^{-2} B] + \ldots \right)
\eeq

Finally, we note that the time component $A_0$ of the gauge connection is not an independent variable: The constraint (\ref{constraint}) can be written as
\beq
\frac{1}{2 \lambda^2} (-D^2 A_0 + \nabla_i \nabla_0 A^i + i [\nabla_0 A_i, A^i]) + \frac{k}{2 \pi} B = 0 
\eeq
and solved for $A_0$ in terms of the spatial components $A_i$:
\beq \label{A0}
A_0 = D^{-2} \left(\nabla_i \nabla_0 A^i + i [\nabla_0 A_i, A^i] \right) + \frac{\lambda^2 k}{\pi} D^{-2} B.
\eeq

To summarize, once a choice of gauge has been made, the phase space can be parametrized by the scalar fields $E_\perp$ and $B$. The quantities $A_i$, $E_i$ and $E_\parallel$ are then not independent variables, but are defined in terms of $E_\perp$ and $B$ by the gauge condition (\ref{gaugechoice}), the decomposition (\ref{parallel_transverse}) and equation (\ref{E_parallel}) respectively. Also the temporal component $A_0$ of the gauge connection is a dependent variable given by the expression (\ref{A0}).

\section{The symplectic structure}
The most familiar route to determining the symplectic structure of a theory with constraints, like the Yang-Mills-Chern-Simons theory, is to identify a set of configuration space coordinates like $A_\mu$ and determine the corresponding canonical momenta $\Pi^\mu = \frac{\partial {\cal L}}{\partial \dot{A_\mu}}$ from the Lagrangian density ${\cal L}$ in (\ref{Lagrangian density}). However, the canonical Poisson bracket $\{\Pi^\mu (x), A_\nu (x^\prime) \}_\mathrm{PB} = \delta^\mu_\nu \delta (x- x^\prime)$ between the coordinates and the momenta would not be consistent because of the constraints on the latter. This can be dealt with by imposing additional constraints (i.e. a gauge condition $\phi_A = 0$), chosen so that the Poisson bracket is non-degenerate on the space spanned by the constraints. There is then an explicit procedure for constructing the Dirac modification $\{ . \, , .\}$ of the Poisson bracket $\{ . \, , .\}_\mathrm{PB}$, so that it will be consistent with the constraints and fullfil all the other requirements on the symplectic structure on the phase space \cite{Dirac}.

This procedure is not obvious to implement in the present case because of the implicit definition (\ref{parallel_transverse}) of our coordinate $E_\perp$. We will therefore follow a different strategy: The Dirac bracket $\{ . \, , .\}$ can be described by the bilinear functionals $e$, $f$ and $g$ defined by
\bea \label{functionals}
e [S; T] & = & \left\{ \integral \Tr (S B), \integral \Tr (T B) \right\} \cr
f [S; T] & = & \left\{ \integral \Tr (S E_\perp), \integral \Tr (T B) \right\} \cr
g [S; T] & = & \left\{ \integral \Tr (S E_\perp), \integral \Tr (T E_\perp) \right\} .
\eea
for any c-number Lie algebra-valued scalar fields $S$ and $T$. These functionals will be uniquely determined by the requirement that the Dirac bracket together with the Hamiltonian (\ref{Hamiltonian}) reproduces the correct equations of motion (\ref{HamiltonianEOM}). 

To begin with, we note that the functional $e$ must vanish identically:\footnote{A hypothetical theory with a non-vanishing equal-time Dirac bracket between the magnetic field at different spatial points would be very different from Yang-Mills theory. The result that $e$ vanishes identically would also come out of the Dirac procedure, since the Poisson bracket between $A_i$ and a gauge condition formulated purely in terms of the connection vanishes. The Dirac bracket thus agrees with the Poisson bracket and vanishes when evaluted between functionals of $A_i$ only.} 
\beq
e [S; T] = 0 .
\eeq

Next, we determine the functional $f$. We start by rewriting
\beq \label{VE}
\integral \Tr (V^i E_i) = \integral \Tr \left({\cal W}_V E_\perp - \frac{\lambda^2 k}{\pi} D_i V^i D^{-2} B \right) ,
\eeq
where 
\beq \label{calW}
{\cal W}_V = \he^{i j} D_i V_j - i [D^{-2} D_i V^i, B]
\eeq
for any c-number\footnote{i.e. with identically vanishing Dirac brackets.} Lie algebra-valued vector field $V^i$. Thus
\beq \label{EiB}
\left\{ \integral \Tr (V^i E_i), \integral \Tr (T B) \right\} = f [{\cal W}_V; T] .
\eeq
If we now apply this formula with $V_i = E_i$ (although this is not a c-number field), we get
\beq
\left\{ \frac{1}{2} \integral \Tr (E^i E_i), \integral \Tr (T B) \right\} = f \left[{\cal W}_E; T \right] ,
\eeq
where the prefactor $\frac{1}{2}$ in the first argument of the Dirac bracket compensates for the integrand being of order $2$ in $E_i$. 
Similarly,
\bea
\left\{ \frac{1}{2} \integral \Tr (B B), \integral \Tr (T B) \right\} & = & e [B; T] \cr
& = & 0 .
\eea
Thus
\beq
\left\{H, \integral \Tr (T B) \right\} = f \left[{\cal W}_E; T \right] ,
\eeq
where $H$ is the Hamiltonian. Since
\bea
{\cal W}_E & = & \he^{i j} D_i E_j - i [D^{-2} D_i E^i, B] \cr
& = & \he^{i j} D_i E_j + \frac{\lambda^2 k}{\pi} i [B, D^{-2} B] ,
\eea
where we have used the constraint (\ref{constraint}) in the last step, we see that
\beq
\left\{H, \integral \Tr (T B) \right\} = \integral \Tr \left((\he^{i j} D_i E_j + i [B, \Lambda]) S^\prime \right) 
\eeq
as required, provided that the functional $f$ is given by
\beq 
f [S; T] = \lambda^2 \integral \Tr (S T)
\eeq
and the gauge parameter $\Lambda$ is given by
\beq \label{Lambda}
\Lambda = \frac{\lambda^2 k}{\pi} D^{-2} B .
\eeq
Such a gauge transformation is thus an unavoidable part of the structure for a theory with a Chern-Simons term. We note that this $\Lambda$ is part of the expression (\ref{A0}) for the temporal component $A_0$ of the gauge connection, which enters in the definition of the gauge covariant temporal derivative.

Finally, the functional $g$ is uniquely determined by the requirement that the symplectic structure obey the Jacobi identity, i.e. that it is given by a closed two-form on the phase space: In view of our earlier results for the functionals $e$ and $f$, the identities
\bea
0 & = & \left\{ \left\{ \integral \Tr (S B), \integral \Tr (S^\prime B) \right\}, \integral \Tr (S^{\prime \prime} B) \right\} \cr
& & + \left\{ \left\{ \integral \Tr (S^\prime B), \integral \Tr (S^{\prime \prime} B) \right\}, \integral \Tr (S B) \right\} \cr
& & + \left\{ \left\{ \integral \Tr (S^{\prime \prime} B), \integral \Tr (S B) \right\}, \integral \Tr (S^\prime B) \right\} \cr
\eea
and
\bea
0 & = & \left\{ \left\{ \integral \Tr (S B), \integral \Tr (S^\prime B) \right\}, \integral \Tr (S^{\prime \prime} E_\perp) \right\} \cr
& & + \left\{ \left\{ \integral \Tr (S^\prime B), \integral \Tr (S^{\prime \prime} E_\perp) \right\}, \integral \Tr (S B) \right\} \cr
& & + \left\{ \left\{ \integral \Tr (S^{\prime \prime} E_\perp), \integral \Tr (S B) \right\}, \integral \Tr (S^\prime B) \right\} \cr
\eea
are automatically fulfilled for any c-number fields $S$, $S^\prime$ and $S^{\prime \prime}$. The identity
\bea
0 & = & \left\{ \left\{ \integral \Tr (S B), \integral \Tr (S^\prime E_\perp) \right\}, \integral \Tr (S^{\prime \prime} E_\perp) \right\} \cr
& & + \left\{ \left\{ \integral \Tr (S^\prime E_\perp), \integral \Tr (S^{\prime \prime} E_\perp) \right\}, \integral \Tr (S B) \right\} \cr
& & + \left\{ \left\{ \integral \Tr (S^{\prime \prime} E_\perp), \integral \Tr (S B) \right\}, \integral \Tr (S^\prime E_\perp) \right\} \cr
& = & \left\{ g [S^\prime; S^{\prime \prime}], \integral \Tr (S B) \right\}
\eea
requires the functional $g$ to be independent of $E_\perp$. By invariance under the gauge group we must have
\beq
\left\{ g [S^\prime; S^{\prime \prime}], \integral \Tr (S E_\perp) \right\} = \integral \Tr \left([S^\prime, S^{\prime \prime}] S \right) \varphi (B) ,
\eeq
for some gauge invariant function $\varphi$ of $B$. The last Jacobi identity then reads
\bea
0 & = & \left\{ \left\{ \integral \Tr (S E_\perp), \integral \Tr (S^\prime E_\perp) \right\}, \integral \Tr (S^{\prime \prime} E_\perp) \right\} \cr
& & + \left\{ \left\{ \integral \Tr (S^\prime E_\perp), \integral \Tr (S^{\prime \prime} E_\perp) \right\}, \integral \Tr (S E_\perp) \right\} \cr
& & + \left\{ \left\{ \integral \Tr (S^{\prime \prime} E_\perp), \integral \Tr (S E_\perp) \right\}, \integral \Tr (S^\prime E_\perp) \right\} \cr
& = & \left\{ g[S; S^\prime], \integral \Tr (S^{\prime \prime} E_\perp) \right\} \cr
& & + \left\{ g[S^\prime; S^{\prime \prime}], \integral \Tr (S E_\perp) \right\} \cr
& & + \left\{ g[S^{\prime \prime}; S], \integral \Tr (S^\prime E_\perp) \right\} \cr
& = & 3 \integral \Tr \left([S^\prime, S^{\prime \prime}] S \right) \varphi (B) ,
\eea
from which follows that $\varphi$ must vanish identically so that
\beq
g [S; T] = 0 .
\eeq

To summarize, to reproduce the correct equations of motion, the symplectic structure on the phase space must thus be given by
\bea \label{symplectic}
\left\{ \integral \Tr (S B), \integral \Tr (T B) \right\} & = & 0 \cr
\left\{ \integral \Tr (S E_\perp), \integral \Tr (T B) \right\} & = & \lambda^2 \integral \Tr (S T) \cr
\left\{ \integral \Tr (S E_\perp), \integral \Tr (T E_\perp) \right\} & = & 0  
\eea
for arbitrary c-number Lie algebra-valued scalar fields $S$ and $T$.

We will also need the Poisson-Dirac bracket for quantities involving covariant derivatives. These can be determined by choosing a gauge and expressing the connection $A_i$ in terms of the magnetic field strength $B$. An alternative gauge invariant argument is to use the relation 
\beq
\delta B = \he^{i j} D_i \delta A_j
\eeq 
for the variation of $B$ induced by an arbitrary variation $\delta A_j$. It follows that the symplectic structure (\ref{symplectic}) can be alternatively expressed as
\bea
\left\{ \integral \Tr (U^i A_i ), \integral \Tr (V^j A_j) \right\} & = & 0 \cr
\left\{ \integral \Tr (S E_\perp), \integral \Tr (V^j A_j) \right\} & = & \integral \he^{i j} \Tr (V_j D_i D^{-2} S) \cr
\left\{ \integral \Tr (S E_\perp), \integral \Tr (T E_\perp) \right\} & = & 0 
\eea
for arbitrary c-number Lie algebra-valued scalar fields $S$ and $T$ and vector fields $U^i$ and $V^i$. 

The Poisson-Dirac brackets involving the electric field strength vector $E_i$ are of particular interest. To investigate these, we start by noting that for any quantity $X$ we have
\bea
\left\{ \integral \Tr (S E_\perp), D_i X \right\} & = & D_i \left\{ \integral \Tr (S E_\perp), X \right\} \cr
& & - \lambda^2 i \he_{i j} [X, D^j D^{-2} S] .
\eea
Applying this to $X = D^i Y$ for an arbitrary $Y$ gives
\bea
\left\{ \integral \Tr (S E_\perp), D^2 Y \right\} & = & D^2 \left\{ \integral \Tr (S E_\perp), Y \right\} \cr
& & - 2 \lambda^2 i \he^{i j} [D_i Y, D_j D^{-2} S] \cr
& & + \lambda^2 [Y, [D^{-2} S, B]] .
\eea
An if finally $Y = D^{-2} Z$ for an arbitrary $Z$, this reads
\bea
\left\{ \integral \Tr (S E_\perp), Z \right\} & = & D^2 \left\{ \integral \Tr (S E_\perp), D^{-2} Z \right\} \cr
& & - 2 \lambda^2 i \he^{i j} [D_i D^{-2} Z, D_j D^{-2} S] \cr
& & + \lambda^2 [D^{-2} Z, [D^{-2} S, B]] ,
\eea
so that
\bea
\left\{ \integral \Tr (S E_\perp), D^{-2} Z \right\} & = & D^{-2} \left\{ \integral \Tr (S E_\perp), Z \right\}  \cr
& & + 2 \lambda^2 i \he^{i j} D^{-2} [D_i D^{-2} Z, D_j D^{-2} S] \cr
& & - \lambda^2 D^{-2} [D^{-2} Z, [D^{-2} S, B]] .
\eea
We now define the auxiliary functional $\gamma$ by
\beq
\gamma [S; V]  = \left\{ \integral \Tr (S E_\perp), \integral \Tr (V^i E_i) \right\}
\eeq
for arbitrary c-number Lie algebra-valued scalar and vector fields $S$ and $V$. It then follows from (\ref{VE}) and (\ref{calW}) and the analogous formulas with $V^i$ replaced by $U^i$ that
\beq
\left\{ \integral \Tr (U^i E_i), \integral \Tr (V^j E_j) \right\} = \gamma [{\cal W}_U; V] - \gamma [{\cal W}_V; U] .
\eeq
Applying this with $U^i = E^i$ gives with the same reasoning as above that
\beq
\left\{ \frac{1}{2} \integral \Tr (E^i E_i), \integral \Tr (V^j E_j) \right\} = \gamma [{\cal W}_E; V] - \gamma [{\cal W}_V; E] .
\eeq
We also have that
\bea
\left\{ \frac{1}{2} \integral \Tr (B B) \right. & \!\!\!\!\!\!\!\!\!\!\!\! , & \!\!\!\!\!\!\!\!\!\!\!\! \left. \integral \Tr (V^j E_j) \right\} \cr
& = & - f [{\cal W}_V; B] \cr
& = & - \lambda^2 \integral \Tr \left((\he^{i j} D_i V_j - i [D^{-2} D_i V^i, B]) B \right) \cr
& = & - \lambda^2 \integral \Tr \left(\he^{i j} V_i D_j B \right) ,
\eea
so that
\bea
\left\{ H, \integral \Tr (V^j E_j) \right\} & = & - \integral \Tr \left(\he^{i j} V_i D_j B \right) \cr
& & + \frac{1}{\lambda^2} \left(\gamma [{\cal W}_E; V] - \gamma [{\cal W}_V; E] \right) .
\eea
In view of the second equation in (\ref{HamiltonianEOM}) we thus find that $\gamma$ must obey
\bea \label{condition}
\gamma [{\cal W}_E; V] - \gamma [{\cal W}_V; E] & = & - \frac{\lambda^4 k}{\pi} \integral \Tr \left(\he^{i j} E_i V_j \right) \cr
& & + i \lambda^2 \integral \Tr \left([ E^i, \Lambda] V_i \right) ,
\eea
with the gauge parameter $\Lambda$ again given by equation (\ref{Lambda}). But, as we have seen above, this equation is not really needed to determine the unique symplectic structure on the phase space.

The above quantities involving the functional $\gamma$ may in principle be explicitly computed, but the expressions are lengthy. Here we content ourselves by giving the results to leading order in the magnetic field strength $B$:\footnote{We will think of these quantities as expressions in terms of $B$ and the gauge covariant operators $D_i$, $D^2$ and $D^{-2}$ rather than the ordinary (metric covariant) operators $\nabla_i$, $\nabla^2$ and $\nabla^{-2}$.}
\bea
\gamma [S; V] & = & \integral \Tr \Bigl(- \frac{\lambda^4 k}{\pi} D_i V^i D^{-2} S \cr
&& - \lambda^2 i ([V_k, D^k D^{-2} S] + [D^{-2} D_k V^k, S]) E_\perp \Bigr) \cr
& & + {\cal O} (B) \cr
\gamma [{\cal W}_U; V] - \gamma [{\cal W}_V; U] & = & \integral \he^{i j} \Tr \Bigl(- \frac{\lambda^4 k}{\pi} U_i V_j \cr
& & + \lambda^2 i D_i ([D^{-2} D_k U^k, V_j] - [D^{-2} D_k V^k, U_j]) E_\perp \Bigr) \cr
& & + {\cal O} (B) \cr
& = &  \integral \Tr \Bigl(- \he^{i j} \frac{\lambda^4 k}{\pi} U_i V_j \cr
& & + \lambda^2 i ([D^{-2} D_k U^k, V_j] - [D^{-2} D_k V^k, U_j]) E^j \Bigr) \cr
& & + {\cal O} (B) .
\eea
Putting $U_i = E_i$ in the last expression, we recognize
\beq
\Tr \left(- \he^{i j} \frac{\lambda^4 k}{\pi} U_i V_j \right) = \Tr \left(- \he^{i j} \frac{\lambda^4 k}{\pi} E_i V_j \right)
\eeq
and
\beq
\lambda^2 i \Tr \Bigl([D^{-2} D_k U^k, V_j] E^j \Bigr) = \lambda^2 \Tr \Bigl(i [E_j, \Lambda] V^j \Bigr)
\eeq
as the two terms in the right hand side of (\ref{condition}), while
\beq
\Tr \Bigl([D^{-2} D_k V^k, U_j]) E^j \Bigr) = \Tr \Bigl([D^{-2} D_k V^k, E_j]) E^j \Bigr) 
\eeq
vanishes identically. 

\section{The Hamiltonian}
By using equations (\ref{parallel_transverse}), (\ref{newconstraint}) and (\ref{E_parallel}) and partial integrations, the Hamiltonian $H$ can be written in terms of the variables $E_\perp$ and $B$ as
\bea
H & = & \frac{1}{2 \lambda^2} \integral \Tr (B B + E_i E^i) \cr
& = & \frac{1}{2 \lambda^2} \integral \Tr \Bigl(B B - E_\perp D^2 E_\perp \cr
& & \;\;\;\;\;\;\;\;\;\;\;\;\;\;\;\;\;\;\;\;\;\;\;\;\;\;\;\;\;\; - [E_\perp, B] D^{-2} [E_\perp, B] - \frac{\lambda^4 k^2}{\pi^2} B D^{-2} B \Bigr) .
\eea

This is non-polynomial in $B$, which also enters implicitly through the connection $A_i$ in the covariant derivative $D_i$. To expand the Hamiltonian as a power series in the coupling constant $\lambda$, we introduce the canonically normalized variables $b$ and $e$ through
\bea
B & = & \lambda b \cr
E & = & \lambda e .
\eea
Just like $B$ and $E_\perp$, the rescaled variables $b$ and $e$ are Darboux coordinates on the phase space:
\bea
\left\{ \integral \Tr (S b), \integral (S^\prime b) \right\} & = & 0 \cr
\left\{ \integral \Tr (S e), \integral (S^\prime b) \right\} & = & \integral \Tr (S S^\prime) \cr
\left\{ \integral \Tr (S e), \integral (S^\prime e) \right\} & = & 0
\eea
for any  c-number Lie algebra-valued scalar fields $S$ and $S^\prime$.

We can now express the Hamiltonian as a power series in $\lambda$. Choosing e.g. Coulomb gauge we have from equation (\ref{Coulomb}) that
\beq
A_i = \he_i{}^j \nabla_j \left(- \lambda \nabla^{-2} b - \frac{i \lambda^2}{2} \he^{k l} \nabla^{-2} [\nabla_k \nabla^{-2} b, \nabla_l \nabla^{-2} b]+ {\cal O} (\lambda^3) \right) .
\eeq
This gives
\bea
H & = & \frac{1}{2} \integral \Tr \Bigl(b b - e \nabla^2 e \Bigr) \cr
& & + i \lambda \integral \he^{i j} \Tr \Bigl( [\nabla_i e, \nabla_j e] \nabla^{-2} b \Bigr) \cr
& & + \frac{\lambda^2}{2} \integral \Tr \Bigl(- [e, b] \nabla^{-2} [e, b] - [e, \nabla_i \nabla^{-2} b] [e, \nabla^i \nabla^{-2} b] \cr
& & \;\;\;\;\;\;\;\;\;\;\;\;\;\;\;\;\;\;\;\;\;\;\;\;\;\;\; - 2 [\nabla_i e, \nabla_j e] \nabla^{-2} [\nabla^i \nabla^{-2} b, \nabla^j \nabla^{-2} b] \Bigr) \cr
& & + {\cal O} (\lambda^3) .
\eea
The Chern-Simons level $k$ will only affect terms of order $\lambda^4$ or higher.

The abelian theory is of course free and has the bilinear harmonic oscillator Hamiltonian
\beq
H = \frac{1}{2} \integral \Tr \left(b (1 - \frac{\lambda^4 k^2}{\pi^2} \nabla^{-2}) b + e (- \nabla^2) e \right) .
\eeq
Excitations which are eigenfunctions to the spatial Laplacian $p^2 = - \nabla^2$ with eigenvalue $p^2$ then have energy
\bea
E & = & \sqrt{(1 - \frac{\lambda^4 k^2}{\pi^2} \nabla^{-2}) (-\nabla^2)} \cr
& = & \sqrt{-\nabla^2 + \frac{\lambda^4 k^2}{\pi^2}} \cr
& = & \sqrt{p^2 + m^2} ,
\eea
where we have introduced
\beq
m = \frac{\lambda^2 k}{\pi} .
\eeq
This is in agreement with well-known results about the mass-gap of Yang-Mills-Chern-Simons theory.

\end{document}